\documentclass[twocolumn,amsmath,amssymb,aps,prl,10pt,lettersize,superscriptaddress,showpacs]{revtex4-1}

\usepackage[dvips]{graphicx} 
\usepackage{braket}
\usepackage[usenames]{color}
\usepackage{ulem}
\graphicspath{{./Figure/}}
\renewcommand{\prl}{Phys.\ Rev.\ Lett.\ }

\newcommand{\PRL}[2]{\prl \textbf{#1}, #2 }

\newcommand{\articletitle}[1]{}

\def\Gakushuin{Faculty of Science, Gakushuin University, 
Mejiro, Tokyo 171-8588,Japan}
\def\UTokyo{Research Center for Advanced Science and Technology (RCAST), The University of Tokyo, \\
Meguro–ku,Tokyo 153–8904, Japan}

\begin{document}

\title{Stable magnetic levitation of soft ferromagnets for macroscopic quantum mechanics}

\author{Maria Fuwa}
\email{maria.fuwa.uni@gmail.com}
\affiliation{\Gakushuin}
\affiliation{\UTokyo}
\date{\today}

\begin{abstract} 

We propose a system for passive magnetic levitation and three-dimensional harmonic trapping of soft ferromagnets. 
Our protocol utilizes the magnetic field gradient for vertical trapping, and the finite size effect of the Meissner effect for horizontal trapping. 
We provide numerical and analytical estimations of possible mechanical dissipations to show that our system allows high mechanical Q-factors above $ Q > 10^8 $, and quantum control of the levitated object is within reach of current technologies. 
The utilization of soft ferromagnet’s internal collective spin excitation may allow quantum mechanical phenomena with particles as large as the sub-millimeter-scale. 

\end{abstract}

\maketitle



\section{I. Introduction}

Levitation of rigid bodies in free space provides an isolated system that circumvents mechanical clamping losses, dissipation, and decoherence, which are usually a limiting factor in mechanical systems~\cite{Gonzalez-BallesteroRev, MaComPhys, QinOptica}.
A method that allows friction-less passively stable levitation without heating or perturbations is magnetic levitation involving diamagnetic objects~\cite{CirioPRL, RomeroPRL12}. 
Magnetic levitation has enabled experimental levitation of hard magnets, diamagnets and superconductors of various size from nanoparticles~\cite{DigiacomoNanoMat}, micrometer-sized spheres~\cite{LengPRAp, WangPRAp, LatorreIEEE, GieselerPRL, VinantePRap, HoferArxiv, LewandowskiPRAp, LatorrePRAp}, millimeter-scale objects~\cite{SchuckSciA, RautIEEE, NakajimaPRA, JiangAPL, XiongPRAp} to centimeter-sized spheres~\cite{ChenAS, RomagnoliArXiv} with mechanical $Q$-factors ranging from $ Q \sim 10^3 - 10^7 $. 
Larger objects have the advantage of better crystal properties, reduced dissipation from heat, gas or acoustic damping due to lower surface-to-volume ratios, and reduced decoherence from vibration, magnetic fluctuations or other force fluctuations due to large mass, albeit subject to more eddy current damping. 
This ability to levitate orders of more massive objects compared to optical tweezers or Paul traps opens a potential for ultra-precise acceleration sensors~\cite{Prat-CampsPRL},  gravimeters~\cite{Goodkind} as well as magnetometers~\cite{KimballPRL}. 

However, trapping and cooling of massive systems larger than micrometer-scales becomes more difficult as the size of the particle increases. 
This is due to the fact that its interaction with control fields used to cool, manipulate or readout the trapped object depends on the single excitation coupling strength $ g = \eta x_{\mathrm{zpf}} $, where $ \eta $ is the coupling strength to the particle's position, and $ x_{\mathrm{zpf}} \sim \sqrt{\hbar / 2 m \omega } $ is the zero point fluctuation with $\omega$ being the center of mass oscillation frequency. 
Since the mass $m = \rho V$ is proportional to the volume $V$, the lighter or smaller the particle, the easier it is to cool to the ground state. 

This can be overcome by utilizing the internal spin degrees of freedom of a levitated soft ferromagnetic (SF) oscillator~\cite{BallesteroPRL, BallesteroPRB, SebersonJOSAB}. 
In this case, the reduction in coupling with increasing mass can be compensated by the increasing coupling of collective spin excitation and the control microwave field. 
Thus the single excitation coupling between the motion and control field becomes size independent, and it is possible to cool the center of mass motion of a SF regardless of its size~\cite{KaniPRLcool}. 
Furthermore, collective spins of a SF in the motional ground state can also be used 
for magnetometers~\cite{KimballPRL} which may enable probing physics beyond the standard model~\cite{FadeevQST, FadeevPRD}, to quantum computing using its rotational symmetry~\cite{GrimsmoPRX}. 

Here we propose a system to levitate and trap the center of mass of a SF in a three dimensional harmonic potential. 
Our system utilizes the magnetic field gradient to trap vertically, and the finite size effect of the Meissner effect from a superconductor whose size is close to the SF to trap horizontally; 
therefore it retains the advantage of magnetic levitation to levitate massive objects. 
Since this system constitutes of only the SF and a superconductor in an external magnetic field, and a superconductor has no eddy current damping, it is ultimately a very low dissipation system. 
We estimate the $Q$-factor limit of the center of mass motion for a Yttrium Iron Garnet (YIG) sphere with eddy current damping and gas damping. Finally, requirements for the external magnetic field stability is discussed.

\begin{figure}[!t]
\begin{center}
\includegraphics[width= \linewidth, clip]{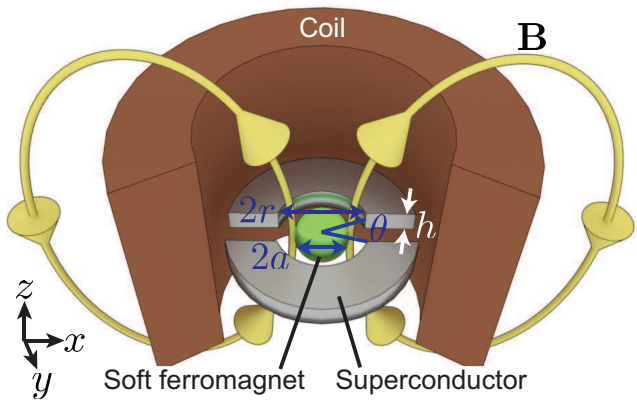}
\caption{Overview: A soft ferromagnet of radius $a$ is placed inside a superconductor with a hole of radius $r$, slit angle $\theta$, height $h$, under a homogeneous external magnetic field $ \mathbf{B}_{\mathrm{ext}} $. Magnetic flux lines are indicated in yellow. Jig is not shown.} 
\label{fig:scheme_overview}
\end{center}
\end{figure}


\section{II. Levitation principal}

\begin{figure}[!t]
\begin{center}
\includegraphics[width= 6 cm, clip]{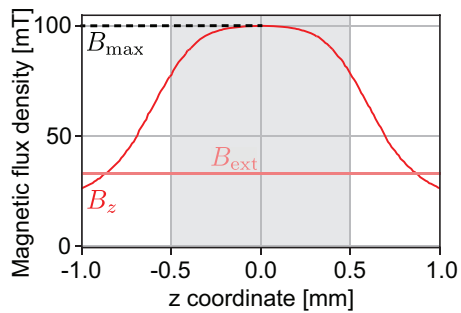}
\caption{Example of magnetic flux density distribution optimized for $B_{\mathrm{max}} = 100$ mT, $2a=0.5$ mm, without the YIG sphere, 
red line: vertical direction $B_z$ ($z$ direction in Fig.~\ref{fig:scheme_overview}) with 
light red line: uniform external magnetic field generated by solenoid coil $B_{\mathrm{ext}}$, 
gray area: inside superconductor hole. 
The origin $(x,y,z)=(0,0,0)$ is defined as the YIG sphere stable levitation point at the center of the superconductor hole. } 
\label{fig:Magfield_Bz}
\end{center}
\end{figure}

An overview of our scheme is given in Fig~\ref{fig:scheme_overview}.
Since Earnshaw's theorem prohibits stable levitation with only static magnetic fields~\cite{Earnshaw}, we use a superconductor disk with a hole and a slit.  
The key difference with levitation of hard magnets is that SF require an external magnetic field $\mathbf{B}$ to saturate the magnetization which is necessary to treat the SF as single domain. 
A superconductor disk in the Meissner state focuses the flux in the direction normal to the disk. 
This can be seen as a ``magnetic flux density lens" that creates a magnetic flux maxima in the center of the hole $\mathbf{B} (0,0,0) \equiv B_{\mathrm{max}} $ (Fig.~\ref{fig:Magfield_Bz}).  
SF can be vertically trapped at the center of this hole, because ferromagnets are attracted to the strongest magnetic field. 
This allows levitation of materials with a permeability of $\mu > 1$ for sufficiently strong $B_z$. 

When the external magnetic field applied on the SF is larger than the saturation magnetization, the magnetization of the SF can be treated as a magnetic dipole. 
In this case, for a SF sphere of radius $ a $, the magnetic levitation force is given by 
\begin{gather}
F_{\mathrm{lev}} 
= M V \frac{d B_z}{dz},
\label{eq:lev_force}
\end{gather}
where $M$ is the magnetization, $V = 4 \pi a^3 /3 $ is the volume, and $B_z$ is the vertical magnetic field density~\cite{MoonWiley}. 
The condition for levitation is for this to exceed the gravitational force $ mg $, with $ m = 4 \pi \rho a^3 /3 $ being the mass, $ \rho $ the density of SF, and $g$ the gravitational acceleration is: 
\begin{gather}
\frac{d B_z}{dz} > \frac{\rho g}{M}.
\label{eq:lev_cond}
\end{gather}
The requirement to apply a magnetic field uniform enough to magnetize the SF into single domain:   
\begin{gather}
\left| \frac{d B_z}{dz}  \right| \ll \frac{ B_{\mathrm{max}} }{a}.
\label{eq:uni_cond}
\end{gather}

The horizontal trapping is achieved using the Meissner effect, in which the image field of the SF sphere exerts a repulsive force towards the center. 
If the SF sphere is small enough compared to the superconductor hole and height, it can be approximated as a magnetic dipole. 
When the size of the superconductor surrounding the SF sphere is comparable to the size of the sphere, finite size effects must be considered. 
The force on the SF needs to be calculated by integrating the Maxwell stress tensor over its surface  
\begin{gather}
\mathbf{F} 
= \int \int \frac{\mathbf{B}_n \cdot \mathbf{B}_n }{2 \, \mu_0} \, d \mathrm{S}. 
\label{eq:horizontal_force}
\end{gather}
where $\mathbf{B}_n$ is the normal component of the sum of external magnetic flux density and image field to the surface~\cite{MoonWiley, HardingNASA}.


\begin{figure*}[!t]
\centering
\includegraphics[width= 0.8\textwidth, clip]{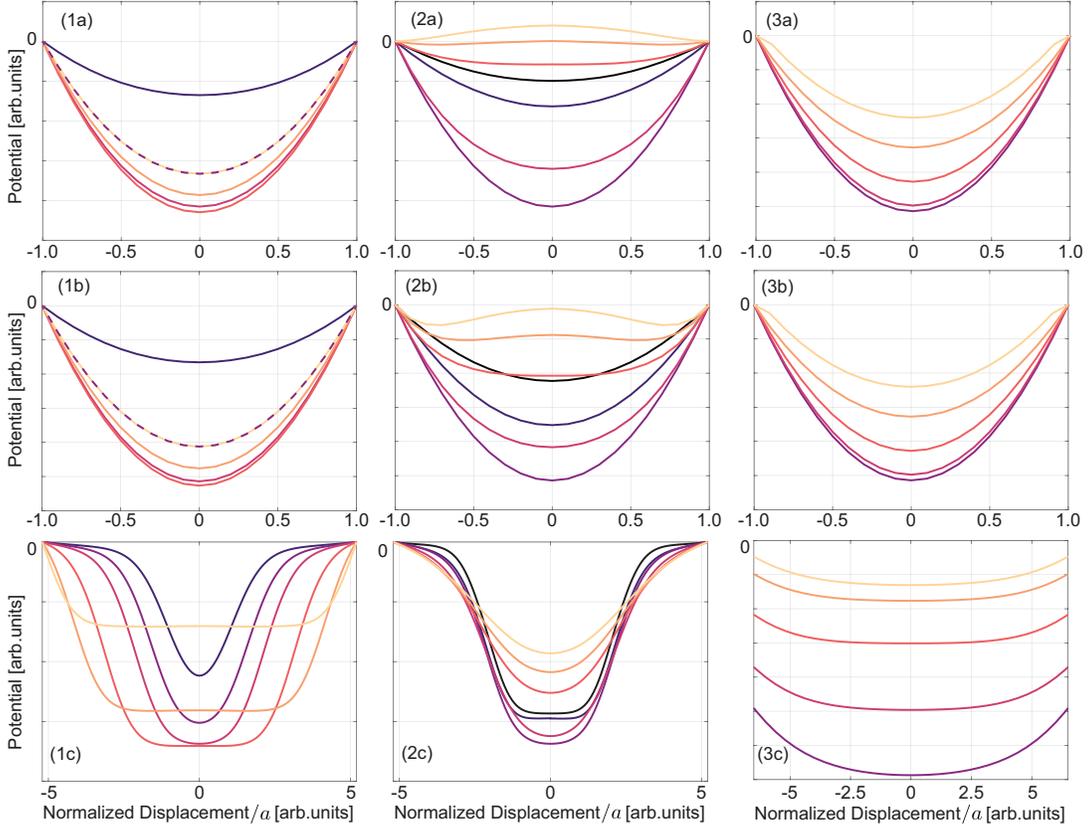}
\caption{Potential energy of YIG sphere dependence for (a) $x$, (b) $y$, (c) $z$ axis, with all displacements normalized by YIG sphere radius $a$.
(1a) - (1c) superconductor height $ h/a = 2, \, 3, \, 4, \, 6, \, 8, \, 10 $ for dark purple to light yellow with $ r/a = 1.4, \, \theta = 10^\circ$. 
(2a) - (2c) superconductor hole radius $ r/a = 1.12, \, 1.2, \, 1.4, \, 1.8, \, 2.2, \, 2.6, \, 3.0 $ for dark purple to light yellow with $ h/a = 4, \, \theta = 10^\circ $.
(3a) - (3c) superconductor slit angle $ \theta = 10^\circ, \, 45^\circ, \, 90^\circ, \, 135^\circ, \, 180^\circ $ for dark purple to light yellow with $ r/a = 1.4, \, h/a = 4$.} 
\label{fig:lev_potential}
\end{figure*}

\section{III. Stable levitation conditions}

While maintaining conditions (\ref{eq:lev_cond}) and (\ref{eq:uni_cond}), we numerically search for the optimal external magnetic field distribution to trap the SF sphere  
in a three dimensional harmonic trap (Fig.~\ref{fig:lev_potential}). 
The height $h$ determines the vertical potential width, while the radius $r$ and slit size $ \theta $ determine the vertical and horizontal potential depth. 

We used a COMSOL software package to simulate the static magnetic field distribution and its magnetic field gradient induced force acting on a Yttrium Iron Garnet (YIG) sphere, by Ampere's Law employing the 3-D finite element method (FEM) (Appendix B). 
We consider YIG as the SF for its low magnetic damping~\cite{SpencerPRL} and high spin density for a ferrimagnet~\cite{ZhangPRL}, which can be exploited for ultrastrong coupling of magnons to microwave cavity modes~\cite{TabuchiPRL}, and ground state cooling of the center of mass motion of a levitated YIG sphere~\cite{KaniPRLcool, BallesteroPRL, SebersonJOSAB}. 
A YIG sphere of permeability $ \mu_\gamma = 32$~\cite{YIGmu-init}, dielectric constant $\epsilon_\gamma = 15 $, density $ \rho_{\mathrm{y}} = 5172 \, \mathrm{kg/m^3} $~\cite{YIGdielectric} is placed at the center of the hole in the superconductor, modeled as a perfect magnetic insulator which fulfills boundary condition $ \mathbf{n} \times \mathbf{A} = 0 $. 
The YIG sphere is stable when there is a resorting force, and the potential is convex downward. 

\begin{figure}[!b]
\begin{center}
\includegraphics[width= \linewidth, clip]{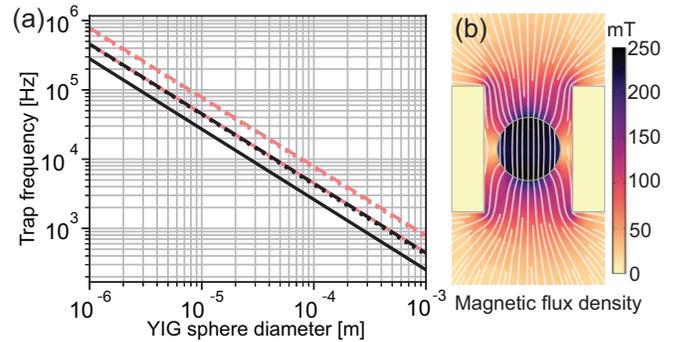} \\
\caption{(a) YIG sphere trap frequency dependence on diameter $2a$ for (light red) $B_{\mathrm{max}} = 100 \, \mathrm{mT}$, (black) $B_{\mathrm{max}} = 1 \, \mathrm{T}$, with solid line: $f_z$, dashed line: $f_y$ respectively.
(b) Magnetic flux distribution of a YIG sphere under an external magnetic field $B_{\mathrm{max}} = 100$ mT when inside a superconductor (yellow rectangle) optimized for harmonic trapping with $r = 1.4 a , \, h = 4a$.}
\label{fig:flux_repulsion}
\end{center}
\end{figure}

The superconductor hole radius $r/a$ determines the magnetic flux concentration in the center of the hole that scales with $ B_{\mathrm{max}} \propto r^{-0.78} $. 
The smaller the hole, the stronger the flux concentration, and stiffer the magnetic spring albeit smaller trapping region. 
The maximum horizontal restoration force is achieved when the radius $r/a$ nearly equals the width of magnetic flux divergence, and any horizontal displacement from the center creates restoring force (Fig.~\ref{fig:flux_repulsion} (b)). 

The superconductor height $ h/a $ determines the homogeneity of the magnetic field in the vertical direction. 
To use the magnetic field gradient for trapping, the height must be short enough so that the uniform magnetic field area is the same size as the YIG sphere.  
The height $ h/a $ has little effect on the strength of magnetic spring $ k_{\mathrm{lev}} $. 

The slit size $ \theta / \pi $ determines flux leakage in the slit direction. 
The ideal condition is when the slit size is infinitely small, allowing magnetic flux to enter into the hole, but does not affect the flux focusing effect nor Meissner effect. 
Thus, the smaller the slit, the stiffer the magnetic spring $ k_{\mathrm{lev}} $, and the more confinement both vertically and horizontally. 
For $ \theta \geq \pi /2 $, there is no trapping in the slit direction ($x$ direction in Fig.~\ref{fig:scheme_overview}).

Taking into consideration the trade-offs above, the superconductor dimensions were optimized by first two-dimensional scanning the hole radius $r/a$ and height $h/a$ with $\theta = 10^\circ$ to find $r/a$ with the deepest horizontal trapping potential. 
Then $h/a$ was optimized to satisfy conditions \eqref{eq:lev_cond} and \eqref{eq:uni_cond}. 
Finally, the limitation on $\theta$ was obtained as the maximum $\theta$ with a convex downward potential. 
We find the YIG sphere can be stably levitated at the center of the superconductor hole when 
$ r/a \sim 1.4 , \, h/a \sim 4 $ and $ \theta \sim 10^{\circ} $, 
for an arbitrary $a$, while satisfying conditions \eqref{eq:lev_cond} and \eqref{eq:uni_cond}.  
For example when $ a = 0.25 $ mm, $ B_{\mathrm{max}} = 100 $ mT, the average 
vertical magnetic field gradient applied over the YIG sphere for $|z| \leq a$ is $ | \overline{ d B_z / dz } | \sim 11.2 $ T/m, which exceeds $ \rho g / M \sim 0.634 $ T/m, while much smaller than $ B_{\mathrm{max}} /a \sim 400 $ T/m.  
The vertical trapping frequency is given by 
\begin{equation}
f_z = \frac{1}{2 \pi} \sqrt{\frac{M}{\rho} \frac{d^2 B_z}{d z^2}}
\propto \frac{ B_{\mathrm{max}} }{a}
\label{eq:freqz}
\end{equation}
in particular $ f_z \sim 0.113 / a $ for 100 mT and $ f_z = 0.217 / a $ for 1 T (Fig.~\ref{fig:flux_repulsion}). 
The horizontal trapping frequencies obey 
$ f_x \sim 1.6 \, f_z$ and $ f_y \sim 1.7 \, f_z $.

\section{IV. Experimental realization} 

Initially, the SF sphere can be positioned at the bottom of the superconductor hole, and a dielectric plate can be placed under the superconductor to support the sphere. 
Spheres can be loaded into the superconductor hole using vacuum tweezers for spheres over $2a \geq 100 \, \mu$m, and micro-manipulators under optical microscopes or electron microscopes for smaller particles~\cite{KomissarenkoNanoMat}. 
Then the SF sphere and superconductor can be cooled simultaneously to a temperature well below its transition temperature $T_{\mathrm{c}}$. 
Finally, the external magnetic field $B_{\mathrm{ext}}$ can be applied with a superconducting solenoid coil to create the desired magnetic field distribution $\mathbf{B}$, lift the SF sphere up, and trap it at the center of the hole. 
Since the critical magnetic field of the superconductor is zero at its transition temperature $H_{\mathrm{c}} ( T_{\mathrm{c}} ) = 0$, zero field cooling is required. 
Thus it is necessary to use superconducting coils and not permanent magnets to generate $B_{\mathrm{ext}}$. 
The use of a silica plate for initial positioning of the SF sphere will cause little eddy current damping and will be negligible after the SF sphere has been trapped (Fig.~\ref{fig:noise_budget}). 

A highly dielectric plate used for initial support can subsequently serve as a microwave resonator that couples to the levitated SF spheres through microwave radiation~\cite{BaeRevIn}. 
For example, cylindrical rutile ($\mathrm{TiO}_2$) resonators with a high dielectric constant of $ \varepsilon \sim 120 $ of diameter 3.6 mm and height 3.0 mm has a fundamental resonant mode ($\mathrm{TE}_{10}$) of $f_0 = 10.3 $ GHz. 
This creates a resonance that induces strong microwave currents on samples up to a few mm above the resonator~\cite{HakkiIEEE}. 
The microwave radiation losses of the rutile resonator can be shielded by the solenoid coil bobbin with lids to enable microwave cavity $Q$-factors in the order of $Q \sim 10^6$~\cite{HashimotoSci}. 
The coupling of this microwave cavity to the internal collective spin excitation of the levitated SF, will enable experimental realization of novel rigid body control protocols~\cite{BallesteroPRL, KaniPRLcool, KaniPRL22}.  

\begin{figure}[!t]
\begin{center}
\includegraphics[width= \linewidth, clip]{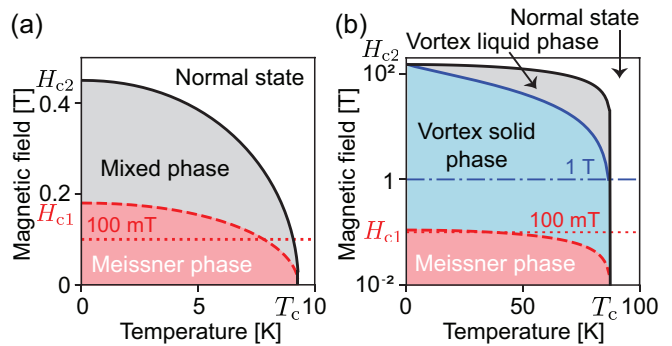}
\caption{A typical H-T phase diagram of (a) Niobium, (b) YBCO plotted with $B_{\mathrm{max}} = 100$ mT (red dotted line), and 1 T (blue dashdot line) respectively.} 
\label{fig:HT_phase}
\end{center}
\end{figure}

Our scheme can be scaled to massive SF spheres as long as the applied external magnetic field is lower than the superconductor's critical magnetic field $ H_{\mathrm{c}} $. 
To apply high enough magnetic fields without destroying the superconductor, type II superconductors can be utilized. 
High purity Niobium (Nb) has a lower critical magnetic field of $ H_{\mathrm{c1}} (0 \, \mathrm{K}) = 180 $ mT~\cite{NbHc1} and upper critical magnetic field of $ H_{\mathrm{c2}} (0 \, \mathrm{K}) = 450 $ mT~\cite{NbHc2} (Fig.~\ref{fig:HT_phase} (a)). 
For higher magnetic fields, high temperature superconductors such as YBCO with the $c$ axis parallel to $B_z$ can be used, with $ H_{\mathrm{c1}} (0 \, \mathrm{K}) = 110 $ mT~\cite{YBCOHc1} and high $ H_{\mathrm{vs}} (0 \, \mathrm{K}) = H_{\mathrm{c2}} (0 \, \mathrm{K}) = 150 $ T~\cite{YBCOHvs} (Fig.~\ref{fig:HT_phase} (b)). 

The Meissner region $ B_{\mathrm{max}} < H_{\mathrm{c1}} $ is ideal region for levitation, when the superconductor exhibits perfect diamagnetism. 
The vortex solid region $ H_{\mathrm{c1}} \leq B_{\mathrm{max}} < H_{\mathrm{vs}} $, when the superconductor is in a mixed state where both superconducting regions and normal regions coexist, but the vortices are pinned in a lattice configuration, can also be used if $ B_{\mathrm{max}} \ll H_{\mathrm{c2}} $ and the superconducting region is dominant. 
This phase exists in hard superconductors such as YBCO with strong flux pinning, but not for soft superconductors such as Nb with weak flux pinning. 
In the vortex liquid region $ H_{\mathrm{vs}} \leq B_{\mathrm{max}} < H_{\mathrm{c2}} $, the highly disordered movement of vortices throughout the superconductor causes substantial energy dissipation, and therefore cannot be used for levitation. 
Here we consider two external magnetic fields of $B_{\mathrm{max}} = 100$ mT where Nb and YBCO 
can be used as superconductor in the Meissner state below $T < 7.8$ K and $T < 36.5$ K respectively, and $B_{\mathrm{max}} = 1$ T where YBCO can be used as a mixed state below $T < 86.6$ K. 

The London penetration depth $\lambda_{\mathrm{L}}$ is a measure of how deeply the magnetic field can penetrate into the superconductor before being expelled.
The London penetration depth of Nb and YBCO are $ \lambda_{\mathrm{L, \, Nb}} (0 \, \mathrm{K}) = 39$ nm~\cite{NbLondon} and $\lambda_{\mathrm{L, \, YBCO}} (0 \, \mathrm{K}) = 100$ nm~\cite{YBCOLondon} and increases with temperature. 
This equivalently rounds the edges of the superconductor by a curvature of approximately $\sim \lambda_{\mathrm{L}}$, leading to an equivalently larger hole diameter. 
Here we consider spheres over $2a > 1 \, \mu$m which are larger than $\lambda_{\mathrm{L}}$, and the magnetic field penetration has little effect (Appendix G).

\begin{figure}[!b]
\begin{center}
\includegraphics[width= \linewidth, clip]{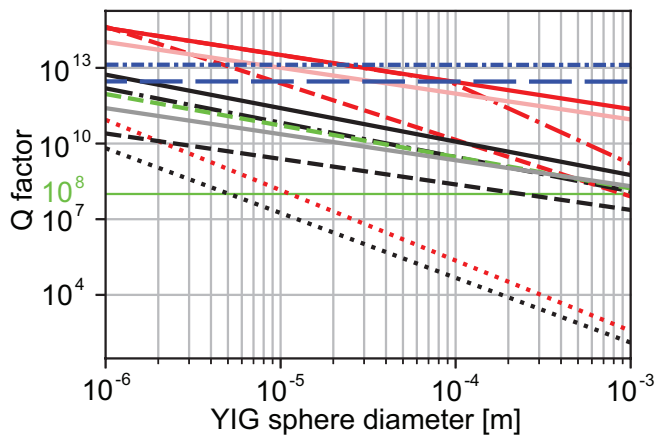}
\caption{Size dependent $Q$-factor limit.
Red line: eddy current damping for $ B_{\mathrm{max}} = 100 \, \mathrm{mT} $, 
Black line: eddy current damping for $ B_{\mathrm{max}} = 1 \, \mathrm{T} $, 
where solid line: YIG internal loss, dashed line: loss from solenoid coil, dotted line: copper plate at $d_{\mathrm{pl}} = 0.1 \, \mathrm{mm}$, dash-dotted line: copper plate at $d_{\mathrm{pl}} = 10 \, \mathrm{mm}$. 
Eddy current damping when dielectric plate at $d_{\mathrm{pl}} = 0 \, \mathrm{mm}$ for 
light red solid line: $ B_{\mathrm{max}} = 100 \, \mathrm{mT} $, 
gray solid line: $ B_{\mathrm{max}} = 1 \, \mathrm{T} $. 
Light green dashed line: eddy current damping from YBCO in mixed state.  
Blue line: gas damping limit for $ P = 10^{-5} \, \mathrm{Pa} $ (dense-dashdot), with squeezed film damping for $ P = 10^{-5} \, \mathrm{Pa} $ (long-dash line).} 
\label{fig:noise_budget}
\end{center}
\end{figure}


\section{V. Dissipation and noise estimate}

Eddy current damping has been the dominant loss in many magneto-mechanical systems, and is in general proportional to the conductivity of the material. 
However, since there is no energy dissipation in the eddy currents in the superconductor surrounding the YIG, our scheme ultimately conserves the magnetic energy and kinetic energy. 

Here we estimate YIG size dependent $Q$-factor limitations from eddy current damping. 
The electromagnetic energy dissipation $ \Delta E_{\mathrm{em}} $ during a single cycle both in the object under consideration and YIG sphere moving in the $z$ direction is calculated using a time varying 3D FEM study in COMSOL multiphysics (Appendix H). 
By comparing this to the kinetic energy of the YIG sphere, the eddy current limited $Q$-factor is 
$ Q_{\mathrm{eddy}} / (2 \pi ) = (1/2 m v^2) \, / \Delta E_{\mathrm{em}} $. 
The YIG internal loss limit is 
$ Q_{\mathrm{eddy, \, 100 \, mT}} > 10^{11} \, ( Q_{\mathrm{eddy, \, 1 \, T}} > 10^8 ) $ 
for $ 2a = $ 1 mm and scales with $ a^{-1.0} \, ( a^{-1.3} ) $. 
The internal loss can be made very small for YIG spheres not only because it is a second order induction effect, but because YIG is an insulator. 

We model the jig as a cylinder larger enough than the levitation system, placed within a distance $ d_{\mathrm{pl}} $ from the surface of the superconductor. 
For a copper plate nearby $ d_{\mathrm{pl}} $ = 0.1 mm, 
$ Q_{\mathrm{eddy, \, 100 \, mT}} \sim 500 \, ( Q_{\mathrm{eddy, \, 1 \, T}} \sim 50 ) $ 
for $ 2a = $ 1 mm, and scales with $ a^{-2.8} \, ( a^{-2.6} ) $ for $B_{\mathrm{max}} = $ 100 mT, 1 T respectively. 
When this plate is moved further away to $ d_{\mathrm{pl}} = 10 \, \mathrm{mm} $, 
$ Q_{\mathrm{eddy, \, 100 \, mT}} > 10^8 \, ( Q_{\mathrm{eddy, \, 1 \, T}} > 10^7 ) $ 
for $ 2a = $ 1 mm, and scales with $ a^{-3.2} \, ( a^{-1.4} ) $. 

The dielectric plate used for initial support of the YIG sphere before levitation placed at $ d_{\mathrm{pl}} = 0 \, \mathrm{mm} $ limits the $Q$-factor by  
$ Q_{\mathrm{eddy, \, 100 \, mT}} > 10^{10} \, ( Q_{\mathrm{eddy, \, 1 \, T}} > 10^7 ) $ 
for $ 2a = $ 1 mm and scales with $ a^{-1.0} \, ( a^{-1.0} ) $. 
This is roughly an order lower than the YIG internal loss limit.  
There was no significant difference between the silica plate and a rutile ($\mathrm{TiO}_2$) microwave cavity. 

The dominant dissipation from the solenoid coil used to create the external magnetic field $B_{\mathrm{ext}}$ is from the bobbin, which can be modeled as a cylinder with hole of diameter $d_{\mathrm{coil}}$. 
For a large enough bore of $d_{\mathrm{coil}} = 40 $ mm, the eddy current damping can be decreased to 
$ Q_{\mathrm{eddy}} > 10^7 $ 
for $ 2a = $ 1 mm and scales with $ a^{-2.2} \, ( a^{-1.0} ) $, which is over an order lower than the YIG internal loss limit. 

Although $B_{\mathrm{max}} = 1$ T is well below $H_{\mathrm{c2}}$ of YBCO, and YBCO is in a mostly-superconducting mixed state, non-zero eddy current damping may arise. 
The penetrated external magnetic field forms normal regions of diameter $~ 2 \xi $ in a triangular lattice configuration of lattice constant $ l_{\mathrm{v}} = 1.075 \sqrt{ \Phi / B_{\mathrm{max}} } \sim 49 $ nm, where $\xi \sim$ 1 nm is the coherence length and $\Phi$ is the magnetic flux quantum~\cite{Tinkham}. 
Here we assume eddy current damping occurs only in these vortices which are in the normal state, and not in the superconducting regions. 
First the volumetric energy loss $ \Delta E_{\mathrm{em}} $ when the whole YBCO bulk is in a normal state is calculated for conductivity $2 \times 10^4$ S/m (at $ T \sim T_{\mathrm{c}} $)~\cite{NamburiSup}. 
This is multiplied by the volumetric ratio of normal region $ \rho_{\mathrm{n}} = 2 \pi \xi^2/( \sqrt{3} \, l_{\mathrm{v}}^2 ) = 1.5 \times 10^{-3} $ to estimate the average volumetric loss from the vortices (Appendix I). 
This gives $ Q_{\mathrm{eddy, \, YBCO}} > 10^{7} $ for $ 2a = $ 1 mm and scales with $ a^{-1.2} $, which is roughly an order lower than the YIG internal loss limit. 
Since the vortex size is much smaller than the skin depth of the ceramic YBCO in a normal state, the eddy current damping may be lower than estimated, which will be experimentally tested elsewhere. 

Thus, $Q_{\mathrm{eddy}} > 10^8 $ is possible for spheres below $2a < 0.2$ mm by using large enough solenoid coils, dielectric jigs instead of copper, and any copper jig placed more than 10 mm away. 

Next, we analytically estimate the gas damping limitation taking in consideration squeezed film damping (SQFD). 
This arises when the gap between the YIG sphere and the superconductor is smaller than the mean free path of free molecules. 
In this case the gas molecules would collide with the oscillator more than if the oscillator were placed in an open space, and the gas damping becomes larger than the vacuum limit. 
The vacuum gas damping limited $Q$-factor is~\cite{WangPRAp}
\begin{gather}
Q_{\mathrm{vac}} = \frac{\pi \rho_{\mathrm{y}}}{6} \sqrt{ \frac{3 k_B T}{m_{\mathrm{g}}}} \frac{a \omega}{P}
\end{gather}
where $\omega$ is the trap angular frequency, Boltzman constant $ k_B = 1.38 \times 10^{-23} \, \mathrm{m^2 \, kg \, s^{-2} \, K^{-1}} $, temperature $T = $ 4 K, mass of gas molecule $ m_{\mathrm{g}} $, and pressure $ P = 10^{-5}$ Pa. 
The gas damping of a sphere inside a cylinder with SQFD is 
\begin{gather}
Q_{\mathrm{sq}} = \frac{16 \rho_{\mathrm{y}}}{3} \sqrt{ \frac{R T}{M_{\mathrm{m}}} } \frac{a^2 (r-a)}{r^2 + 2/3 a^2 - \pi/2 a r} \frac{\omega}{P}
\end{gather}
with gas constant $R = 8.31 \, \mathrm{m^2 \, kg \, s^{-2} \, K^{-1} \, mol^{-1}} $ and molar weight of air $ M_{\mathrm{m}} = 28.966 $ g/mol. 
Since SQFD is strongest against horizontal displacements, we assume $ \omega = 2 \pi f_y $, and ignore the slit for simplicity (Appendix J)~\cite{Bao-RFmodel}. 
Although the SQFD limited $Q$-factor is roughly an order lower than the vacuum limited $Q$-factor, $ Q_{\mathrm{sq}} > 10^{12} $ and is negligible at high enough vacuum $ P = 10^{-5}$ Pa compared to eddy current losses. 

Finally, there are internal magnon losses in YIG due to acoustic damping $ Q_{\mathrm{pn}} \approx 10^5 - 10^7 $ 
and Gilbert damping $ Q_{\mathrm{yig}} \sim 10^4 $. 
However levitated YIG spheres may have internal dissipation limited $Q$-factors as high as $ Q \sim 10^{10} $~\cite{BallesteroPRL}. 

\begin{figure}[!b]
\begin{center}
\includegraphics[width= \linewidth, clip]{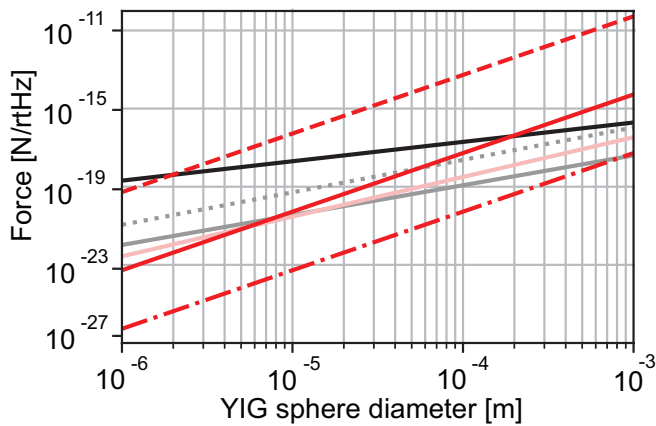}
\caption{Size dependent force noise at trap frequency $f_z$. 
Thermal force 
at 4 K for $Q=10^8$ (black solid line), 4 K when $Q$-factor is YIG internal loss limited for $ B_{\mathrm{max}} = 100 \, \mathrm{mT} $ (gray solid line), $ B_{\mathrm{max}} = 1 \, \mathrm{T} $ (gray dotted line). 
Red line: magnetic force fluctuations for $ \delta B / B = 10^{-6} $ (dashed line), $ 10^{-10} $ (solid line), $ 10^{-13} $ (dashdot line). 
Light red line: seismic noise limit with multistage active controls~\cite{LIGO-budget}. }
\label{fig:force_budget}
\end{center}
\end{figure}

In order to control or readout the quantum motion of the trapped YIG, the force fluctuations must be smaller than the Brownian Force $ 4 k_B T \gamma $, where $\gamma = \omega / Q$ is the mechanical decay rate (Fig.~\ref{fig:force_budget}). 
Vibrations from pulse-tube coolers in cryogen-free cryostats are below 10 Hz and is negligible for the kHz frequencies used to trap $\mu$m YIG spheres~\cite{SchmoranzerCryo}. 
These low frequency vibrations which may limit long term stability can be reduced below room temperature thermal noise level by passive vibration isolation using magnetic vibration dampers and employing flexible heat links in the cryostat~\cite{Fleischer-RevIn}. 
Further vibration reduction can be achieved using multistage active controls to less than the thermal limit at 4 K for $Q > 10^8$~\cite{LIGO-budget}. 

To assess the fluctuations from the external magnetic field, we assume the YIG sphere is in its position of equilibrium, where 
$ F_x = F_y = 0 $, $ F_z = mg $. 
If the magnetic field fluctuation $\delta B$ is small enough compared to the external magnetic flux density $B$, 
$ \delta F / F = 2 \, \delta B / B $ (Appendix K).  
Magnetic shielded rooms can reduce magnetic drifts from stray magnetic fields down to a few fT level~\cite{ILL2019}, which gives a negligible fluctuation of $ \delta B / (100 \, \mathrm{mT}) \sim 10^{-14} $. 
For an external magnetic field created by a solenoid coil driven by a commercial DC current supply, $ \delta B / B = \delta I_{\mathrm{coil}} / I_{\mathrm{coil}} = 10^{-6} $ at DC. 
Since current noise $ \delta I_{\mathrm{coil}} $ decreases with frequency by 1/f due to Flicker noise, relative current noise of $ \delta I_{\mathrm{coil}} / I_{\mathrm{coil}} = 10^{-10} $ is possible for kHz frequencies used to trap $\mu$m YIG spheres using ultra low noise current sources~\cite{SMC11}.  
Thus, for $ 2a < 0.2 $ mm YIG spheres, quantum control and readout of quantum motion is within reach of current technologies. 
An improvement of stability to $ \delta I_{\mathrm{coil}} / I_{\mathrm{coil}} = 10^{-13} $ will allow quantum control of mm scale YIG spheres.


\section{VI. Possible applications}

Taking advantage of its high $Q$-factor, levitated SF may have applications in accelerometers~\cite{TimberlakeAPL}, magnetometers~\cite{KimballPRL}, and gyroscopes~\cite{Prat-CampsPRL}. 
While many magnetic levitation systems utilize SQUID for motion readout, since YIG is an insulator, the motion of YIG spheres can be read out by optical cavities that enable shot noise-limited sensing~\cite{XiongPRAp}. 
Furthermore, utilization of YIG's low internal loss and magneto-crystalline anisotropy may allow magnonic quantum networks~\cite{RusconiPRA} or quantum tunneling~\cite{WernsdorferSci}.

In addition, utilization of coupling internal spin excitations to the levitating rigid body motion can open way to a wide variety of new physics. 
For example, electron spins couple to mechanical rotation through Einstein de Haas physics~\cite{KeshtgarPRB}. 
For nano YIG particles, this can enable spin stabilized magnetic levitation that break Earshaw's theorem~\cite{RusconiPRB16, RusconiPRL17} or fast rotations above 10 GHz~\cite{KaniPRL22} that may enable quantum racket flips~\cite{MaPRL}. 
For micro-meter sized YIG particles, magnetometers with unprecedented scaling~\cite{KimballPRL} may lead to experimental tests of certain axion models~\cite{FadeevQST} or the Lense-Thirring effect on magnetized objects~\cite{FadeevPRD}. 
Since the spin-mechanical coupling is intensive, these proposals hold for arbitrary sized YIG spheres, enabling ground state cooled milli-meter sized YIG spheres with the possibility of motional quantum superpositions. 
Our results may open way to novel spin-optomechanical coupling that enables quantization of rotational modes, quantum superpositions, to gyroscopes, in analogous to NV centers~\cite{DelordNature, HuilleryPRB, PerdriatRev}.


\section{VII. Conclusion}

In conclusion, we have proposed a magnetic levitation and three dimensional harmonic trapping system for arbitrary size soft ferromagnets with variable trapping frequency. 
A soft ferromagnet is trapped in the center of a superconductor hole with a slit under an external magnetic field;
the magnetic field gradient is used for vertical trapping and the finite size effect of the Meissner effect for horizontal trapping. 
The eddy current damping and gas damping has been estimated to enable $Q$-factors over $Q > 10^8 $, and quantum control of spheres below 0.2 mm diameter is within reach of current technologies, by reducing magnetic field fluctuations below $ \delta B / B < 10^{- 10} $. 
In contrary to hard magnet levitation, the internal spin excitations of the soft ferromagnet can be coupled to its rigid body motion, which may enable quantum mechanical phenomena with larger particles.


\section{Acknowledgements}

Maria Fuwa acknowledges financial support from JST, PRESTO (grant number JPMJPR1866) and The Sumitomo Foundation (grant number 210825). 
I also acknowledge the comments and suggestions from Yasunobu Nakamura, Koji Usami, and Kohei Matsuura as well as Nobuyuki Matsumoto for fruitful discussions on squeezed film damping and mechanical dissipation estimation.
\\

\textit{Note added}. Recently, a preprint by Fuwa \textit{et al.} appeared~\cite{YIGlevexp}, demonstrating stable levitation of a sub-milligram, sub-millimeter Yttrium Iron Garnet sphere using the magnetic levitation proposed in this paper.


\section{Appendix}
\section{Appendix A: Flux focusing effect of a superconductor}

Since Earnshaw's theorem prohibits stable levitation with only static magnetic fields~\cite{Earnshaw}, we use a superconductor disk with a hole and a slit. 
The superconductor is zero-field-cooled: it is cooled below its transition temperature before the magnetic field is applied. 
The external magnetic field can be created by a solenoid coil larger enough than the levitation system or a Helmholtz coil, but not by permanent magnets. 
We have modeled the uniform magnetic field $ \mathbf{B} $ by a single solenoid coil larger enough than the YIG sphere. 

Since the magnetic flux inside the superconductor is conserved $ \mathbf{rot} \,  \mathbf{B} = 0 $, a slit is necessary to allow the magnetic flux enter into the hole.
When an infinitely thin superconductor with a hole and infinitely thin slit is inserted, the magnetic flux $ \Phi_{\mathrm{s}} $ of where the superconductor was is concentrated into the hole (Fig.~\ref{fig:Magfield_Bz}). 
Since the magnetic flux inside the superconductor is conserved $ \mathbf{rot} \,  \mathbf{B} = 0 $, 
$ \Phi_{\mathrm{s}} = \pi r^2 \, B_{\mathrm{max}} $. 
In the actual case of finite slit size $ \theta = 10^\circ $, some of the magnetic flux diverges into the slit. 
For a finite height $ h = 10 / 3 \, r $, the magnetic flux is also focused in the direction  of the height ($z$ direction in Fig. 1 of main text). 
The maximum magnetic flux density was calculated from numerical simulations to scale as 
$ B_{\mathrm{max}} \propto B_{\mathrm{ext}} / r^{-0.88} $. 


\section{Appendix B: Three-Dimensional COMSOL model}

A three-dimensional COMSOL model was constructed and used to calculate the magnetic field density spatial distribution, and resulting time-dependent forces and losses acting on the YIG sphere. 
First, the external magnetic field is created by a large enough solenoid coil, which is modeled by a bundle of tiny wires that are tightly wound together and separated by an electrical insulator but are not geometrically resolved. 
This coil is excited by a current flowing only in the directions of the wires.
Then the three-dimensional magnetic field distribution of the whole system is calculated by a stationary study, using the Amp\'ere's Law for the YIG and objects under consideration.  
The force acting on the YIG sphere is calculated by surface integrating the Maxwell tensor 
\begin{gather}
\mathbf{F} 
= \int \int \frac{\mathbf{B}_n \cdot \mathbf{B}_n }{2 \, \mu_0} \, d \mathrm{S}. 
\tag{\ref{eq:horizontal_force}} 
\end{gather}
Trapping potentials (Fig.~\ref{fig:lev_potential}) were calculating by changing the YIG sphere position for varying superconductor sizes. 
Finally, a time-dependent study is used to calculate the dynamics of the YIG sphere to determine the trap frequency and eddy current losses.

\section{Appendix C: Finite element method mesh}

Constructing a fine enough mesh is crucial for finite element calculations, especially in force calculation. 
In our setup, the mesh of the SF sphere itself and the surrounding air within the superconductor hole and slit is critical to get reliable results. 
We simulated the force on a SF sphere with varying minimum allowed mesh element size $l_{\mathrm{mesh}}$ to accurately resolve the small regions between the sphere and superconductor wall. 
A mesh conversion analysis showed that $l_{\mathrm{mesh}} \lessapprox 0.176 \, a $ is sufficient to calculate the force on the SF is accurate for two significant figures.  
The largest mesh element size was $\approx 1.144 \, a$.

We simulated the trap frequencies for a largest mesh element size of 0.286 mm and a slightly varying minimum mesh around $l_{\mathrm{mesh}} \sim 4.4 \, \mu \mathrm{m} $ for a YIG sphere of $a = 0.25 \, \mathrm{mm} $. 
The trapping frequencies were $ f_z =$ 538.75, 538.87, 538.85, 538.85, 538.85, 538.84 Hz, resulting in a mean value of $ f_z = 538.84 \pm 0.05$ Hz. 
This accounts to a 0.009 \% uncertainty in trap frequency which is negligible throughout the paper.

\begin{figure}[!b]
\begin{center}
\includegraphics[width= 6.5 cm, clip]{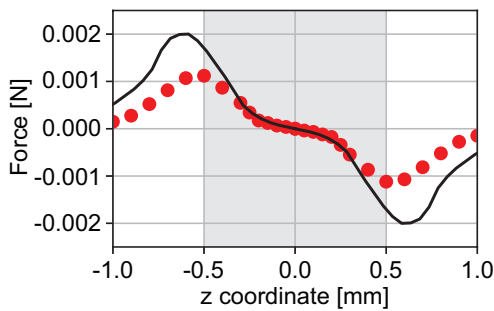}
\caption{Example of vertical magnetic force $F_z$ dependence on YIG sphere position for $B_{\mathrm{max}} = 100$ mT, $2a = 0.5$ mm calculated from 
black line: Eq.(\ref{eq:lev_force}) with $B_z$ from Fig.~\ref{fig:Magfield_Bz}, red dots: Maxwell stress tensor in Eq.(\ref{eq:horizontal_force}), gray area: inside superconductor hole. } 
\label{fig:fz_grad_comparison}
\end{center}
\end{figure}

\section{Appendix D: Magnetic gradient induced force $F_z$} 

To estimate the stable levitation conditions, the vertical magnetic force $F_z$ was approximated as the magnetic gradient induced levitation force on a magnetic point dipole $a \rightarrow 0$ as  
\begin{equation}
F_{\mathrm{lev}} 
= M V \frac{d B_z}{dz}. 
\tag{\ref{eq:lev_force}}
\end{equation} 
To verify whether this equation holds true for finite size spheres, we compare this force with the magnetic force calculated from the Maxwell stress tensor in Eq.(\ref{eq:horizontal_force}) in Fig.~\ref{fig:fz_grad_comparison}. 
Both calculation methods coincide when the YIG sphere is near the center half of the superconductor hole, where the magnetic field gradient is nearly homogeneous over the sphere. 
However, discrepancies arise towards the edges of the superconductor hole, when the magnetic field gradient is nonuniform over the sphere. 
Thus, the magnetic point dipole approximation in Eq.(\ref{eq:lev_force}) is valid when the YIG sphere is near its stable levitation point.

\section{Appendix E: Magnetic point dipole limit}

To show that our horizontal trapping is indeed due to the finite size effect of the image field created inside the superconductor, we compare our results with the limiting case of magnetic point dipole $a \rightarrow 0$. 
The magnetic field density of a magnetic point dipole is 
\begin{gather}
B_{\mathrm{dp}} (\mathrm{r}) 
= \frac{\mu_0}{4 \pi} \frac{m_{\mathrm{dp}}}{|\mathbf{r}|^5} \left( 3 \mathbf{r} (\mathbf{r} \cdot \mathbf{n} ) - \mathbf{n} |\mathbf{r}|^2 \right), 
\tag{E1}
\end{gather}
where $ m_{\mathrm{dp}}$ is the magnetic moment, $\mathbf{r}$ is the position vector, $\mathbf{n}$ is the unit vector in direction of the magnetic moment. 
When this dipole approaches a superconductor in the Meissner state, persist currents in the superconductor are established which produce a magnetic field opposing that of the dipole. 
When the magnetic moment of the dipole is parallel to the surface of the superconductor, the image field reads 
\begin{gather}
B_{\mathrm{dp,im}} (h) = \frac{\mu_0}{\pi h^3} \, m_{\mathrm{dp}}, 
\tag{E2}
\end{gather}
where $h$ is the distance between the dipole and superconductor. 
Thus the force the dipole experiences from a single image field is
\begin{gather}
F_{\mathrm{dp},r} (h) 
= ( m_{\mathrm{dp}} \cdot \nabla ) \, B_{\mathrm{dp,im}} (h)
= \frac{5}{2 \pi} \frac{\mu_0 m_{\mathrm{dp}}^2 }{h^4}. 
\tag{E3}
\end{gather}
When the dipole is placed in between two parallel planes of superconductors of distance $h$ and is displaced by $ \delta r $, the force created by the images fields is 
\begin{align}
&F_{\mathrm{dp},r} (h + \delta r) - F_{\mathrm{dp},r} (h - \delta r) \notag \\
&\quad = \frac{5}{2 \pi} \mu_0 m_{\mathrm{dp}}^2 \left( \frac{1}{(h + \delta r)^4} - \frac{1}{(h - \delta r)^4} \right) \notag \\
&\quad = \frac{5 \mu_0 m_{\mathrm{dp}}^2}{2 \pi} \, \frac{h^2 + (\delta r)^2}{(h^2 - (\delta r)^2)^4} \, h \, \delta r
\tag{E4}
\label{eq_Fr_dipole}
\end{align}
When $\delta r \ll h$ and the sphere is close to the center of the coil, the restoring force is linear to $\delta r$. 
As the sphere approaches the sides of the superconductor $ \delta r \rightarrow h $, the restoring force diverges with $(h - \delta r)^{-4} $. 

\begin{figure}[!t]
\begin{center}
\includegraphics[width= 6.5 cm, clip]{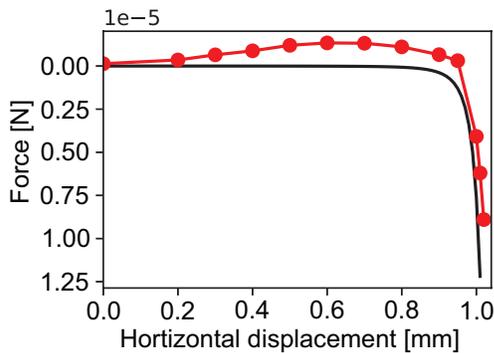}
\caption{Example of the force a YIG sphere of $ a = 50 \, \mu$m experiences from a large enough superconductor of $r = 1.07$ mm, $h = 1$ mm, $B_{\mathrm{max}} = 53.5$ mT. 
(Red) Magnetic force calculated by COMSOL multiphysics, (black) dipole force analytically calculated by Eq.(\ref{eq_Fr_dipole}).} 
\label{fig:mag_dipole_lim}
\end{center}
\end{figure}

We simulate the force acting on a magnetic point dipole created by a YIG sphere that is smaller compared to the size of the superconductor hole and height. 
The restoring force acting on a YIG sphere of $ a = 50 \, \mu$m from a large enough superconductor of $r = 1.07$ mm, $h = 1$ mm under an external magnetic field $ B_{\mathrm{max}} = 53.5$ mT is given in Fig.~\ref{fig:mag_dipole_lim}. 
This shows that our simulations hold true in the dipole limit where the force scales with $(h - \delta r)^{-4} $. 

The discrepancies can be explained by the magnetic gradient induced force as follows. 
If there were zero or a completely uniform external magnetic field, the dipole would experience the force given by Eq.(\ref{eq_Fr_dipole}). 
However, the external magnetic field density in the center of the superconductor is maximum in the vertical direction, and minimum in the horizontal direction. 
This will add a magnetic gradient induced force given by 
\begin{gather}
F_{\mathrm{dp}, grad} = \mathbf{m}_{\mathrm{dp}} \cdot \nabla \mathbf{B}_{\mathrm{ext}} 
\tag{E5}
\end{gather}
where $ \mathbf{m}_{\mathrm{dp}} $ is the magnetic dipole of the YIG sphere and $ \mathbf{B}_{\mathrm{ext}} $ the external magnetic field density. 
Thus, when the YIG sphere is close to the center, it will experience an attractive force towards the sides of the coil. 
As it moves closer to the superconductor, the repulsive force will become dominant.

\section{Appendix F: Magnetization saturation effect}

Here we consider the effect of magnetization saturation on the 
trapping frequency. 
Since YIG has a saturation magnetization of $ M_{\mathrm{sat}} = 196 \, \mathrm{kA/m} \, (\mu_0 M_{\mathrm{sat}} \sim 246 \, \mathrm{mT})$, for $ B_{\mathrm{max}} = 100 $ mT which is below the saturation magnetization, we use the relative permeability $\mu_\gamma $ to calculate constitutive relation $ B = \mu_0 \mu_\gamma H $ used in the Amp\'ere's law. 
Since the trapping frequencies are sufficiently smaller than the ferromagnetic resonance of YIG in the GHz frequencies, where the permeability changes rapidly, we use the initial magnetic permeability at 40 kHz $\mu_\gamma = 32 $~\cite{YIGmu-init} for relative permeability. 
For $ B_{\mathrm{max}} = 1$ T which is well above the saturation magnetization, we use 
the constitutive relation $ B = \mu_0 (H + M_{\mathrm{sat}}) $ in the Amp\'ere's law. 

\begin{figure}[!t]
\begin{center}
\includegraphics[width= 6.5 cm, clip]{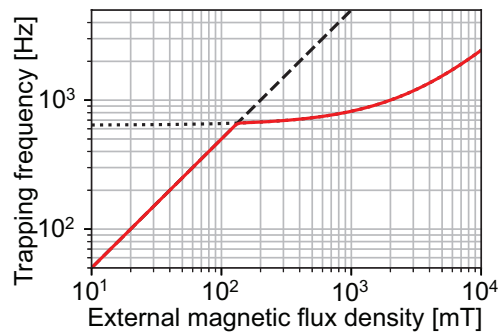}
\caption{Vertical trap frequency $f_z$ dependence on applied external magnetic field $B_{\mathrm{max}}$ for a $a = 0.25$ mm YIG sphere.
Trap frequencies are calculated using the relative permeability (black dashed line) and saturation magnetization (black dotted line), with the lower frequency of the two are showen in red.} 
\label{fig:fz_B_dependence}
\end{center}
\end{figure}

In Fig.~\ref{fig:fz_B_dependence}, we show the trapping frequency dependence on external magnetic field density $B_{\mathrm{max}}$ for a sphere of $a = 0.25 \, \mathrm{mm} $. 
For $ B_{\mathrm{YIG}} \ll \mu_0 M_{\mathrm{sat}} $, the trapping frequency increases proportionally with the external field as 
$ f_z = 5.1 \, B_{\mathrm{max}} $. 
For $ B_{\mathrm{YIG}} \gg \mu_0 M_{\mathrm{sat}} $, while the trapping frequency increases linearly with the external magnetic field
$ f_z = 0.18 \, B_{\mathrm{max}} + 640 $, 
the change is less acute. 
The trapping frequencies coincide at about $B_{\mathrm{max}} \sim 107.4 \, \mathrm{mT}$, which about 43.7 \% of $ \mu_0 M_{\mathrm{sat}} \sim 246 \, \mathrm{mT}$. 

This discrepancy is likely due to the difference in modeling the constitutional equation. 
When using the relative permeability, the material is allowed to respond to the applied field with a changing magnetization; the magnetization of the YIG sphere is slightly non-uniform depending on the time varying external magnetic field distribution. 
When using the magnetization as a prescribed vector, the magnetization of the YIG sphere is aligned with $B_{\mathrm{max}}$ independent of the external field. 
When the force on the YIG sphere is calculated through surface integration of the Maxwell stress tensor (Eq.(\ref{eq:horizontal_force})), this spatial discrepancy of magnetization can cause a discrepancy in electromagnetic force and trap frequency.

\section{Appendix G: London penetration depth}

\begin{figure}[!t]
\begin{center}
\includegraphics[width= \linewidth, clip]{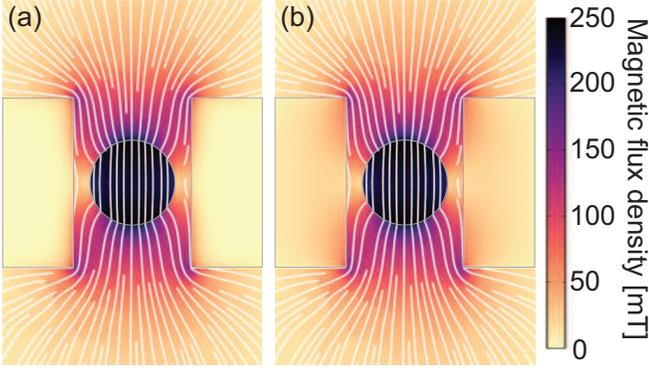}
\caption{Magnetic field distribution considering London penetration for $B_{\mathrm{max}} = 100$ mT,  $\lambda_{\mathrm{L, \, YBCO}} (0 \, \mathrm{K}) = 100$ nm for YIG sphere diameter of (a) $2a=1 \, \mu$m, (b) $2a=100$ nm, with optimal trapping conditions $r=1.4a, \, h=4a$. } 
\label{fig:magf_london}
\end{center}
\end{figure}

The London penetration depth $ \lambda_{\mathrm{L}}$ is a characteristic length in superconductors that describes how far an external magnetic field can penetrate into the material before it is expelled in order to maintain a diamagnetic state with zero resistance. 
We asses this effect on SF spheres of diameter $2a=100$ nm, $1 \, \mu$m. 
First the magnetic field distribution when the superconductor exhibits perfect diamagnetism is calculated by a three dimensional finite element method using COMSOL Multiphysics. 
Subsequently, the magnetic field penetration into the superconductor is calculated using the London equation
\begin{equation}
\nabla \cdot \mathbf{B} = \frac{1}{ \lambda_{\mathrm{L}}^2 } \, \mathbf{B}. 
\tag{G1}
\end{equation} 
where $\lambda_{\mathrm{L}} \equiv \lambda_{\mathrm{L, \, YBCO}} (0 \, \mathrm{K}) = 100$ nm is the London penetration depth of YBCO (Fig~\ref{fig:magf_london}). 
The London penetration is equivalent to rounding the edges of the superconductor by approximately $ \sim\lambda_{\mathrm{L}}$. 

For a $2a=1 \, \mu$m sphere, this penetration length $ \lambda_{\mathrm{L}} $ is smaller than the gap between the superconductor hole and SF sphere $r-a=0.2 \, \mu$m. 
An increase in the effective hole radius to $ \sim (r+ \lambda_{\mathrm{L}}) = 1.6 a $ will have little effect on the trapping potential, as can be seen from Fig.~\ref{fig:lev_potential}.  
For maximum trapping strength, the magnetic flux penetration can be compensated by using a superconductor hole with a smaller diameter of $ \sim 2(r - \lambda_{\mathrm{L}}) = 1.2 \, \mu$m. 

A $2a=100$ nm sphere is the same size as the penetration length $ \lambda_{\mathrm{L}} $, which is larger than the gap between the superconductor hole and SF sphere $r-a=20$ nm. 
An increase in the effective hole radius to $ r+ \lambda_{\mathrm{L}} = 2.4 a $ will result in almost no trapping in the direction normal to the slit ($x$ direction in Fig.~\ref{fig:scheme_overview}). 
This cannot be compensated by using a hole with a smaller diameter. 
Thus the London penetration depth is negligible for the SF spheres of $2a \geq 1 \, \mu$m in consideration.

\begin{figure}[!t]
\begin{center}
\includegraphics[width= 7cm, clip]{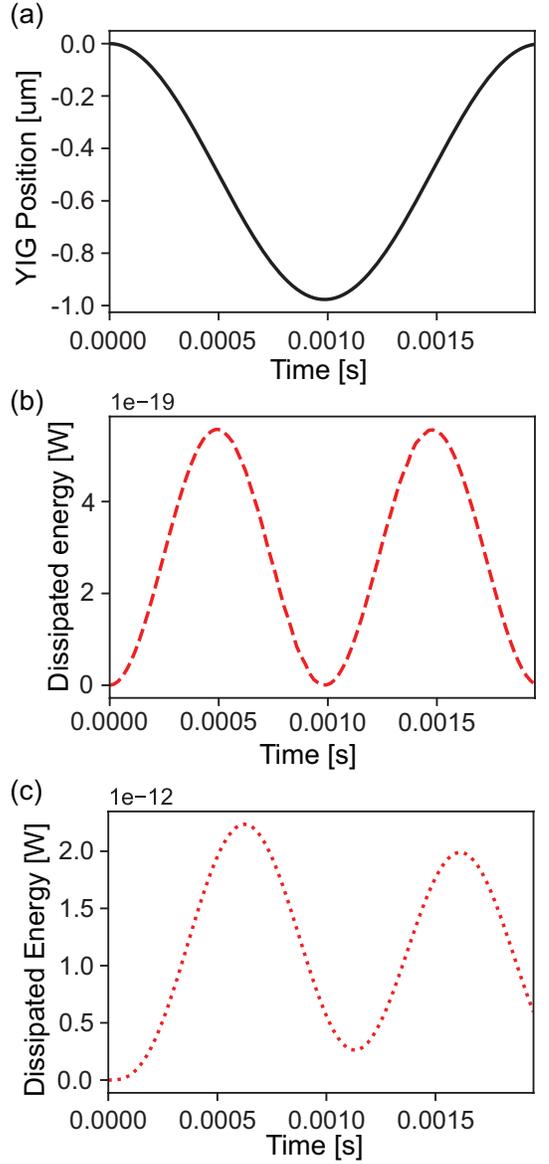}
\caption{Time dependent electro-magnetic energy loss during one cycle for $2 a = 0.5$ mm YIG sphere.
(a) YIG position, 
(b) energy dissipation in YIG sphere,
(c) energy dissipation in a copper plate of distance $d_{\mathrm{pl}} = 0.1$ mm.} 
\label{fig:eddy_current_est}
\end{center}
\end{figure}

\section{Appendix H: Eddy current damping estimation}

When an object moves relative to a conductor, the electromotive force creates a current loop to counteract this motion. 
The energy dissipation from this current flow is known as eddy current damping, and has been the dominant loss in many magneto-mechanical systems. 
The eddy current damping is calculated by the electric energy dissipated per cycle in the YIG and nearby plate using the time-dependent solver in COMSOL.
The magnetic vector potential is calculated from
\begin{gather}
( j \omega \sigma - \omega^2 \varepsilon ) \mathbf{A} + \nabla \left( \frac{1}{\mu} \nabla \times \mathbf{A} \right) = 0
\tag{H1}
\end{gather}
where $\sigma$ is the conductivity, $\varepsilon$ the premittivity, $\mu$ the permeability, $\omega$ the trap angular frequency, and $ \delta = \sqrt{ 2/ \omega \mu \sigma }$ the skin depth. 
For the YIG and silica plate which are insulators, $ \sigma \ll 1 $ and the skin depth $\delta > 1$ km is larger than the objects. 
For the copper plate, we use $ \sigma = 5.998 \times 10^8 $ S/m for low temperatures, 
the skin depth is nearly equal to the YIG diameter $ \delta \sim 2 a $. 
In these cases, the eddy current dissipation can be calculated by volume integration of the Poynting vector
\begin{gather}
P_{\mathrm{eddy}} = \frac{1}{2} ( \mathbf{J}_S \cdot \mathbf{E}* )
\tag{H2}
\end{gather}
where $ \mathbf{J}_S $ is the induced current and $\mathbf{E}$ is the electric field inside the object. 

We use a time-dependent solver in COMSOL to calculate the electrical losses in both the YIG and objects in consideration for a single cylce (Fig.~\ref{fig:eddy_current_est}). 
As the YIG sphere oscillates, the eddy currents induced in both the YIG sphere and surrounding objects causes a displacement dependent dissipation. 
Since the eddy current dissipation in the surrounding objects is a first order induction effect caused by the YIG motion, while the eddy current in the YIG sphere is a second order induction effect caused by the magnetic field change due to eddy current in the plate, the former is larger than the latter. 
In the time dependent study, a cycle is divided into time slots of $ \delta t = 1/ f_z / 100 $ s, and the energy dissipation is calculated for every temporal duration. 
The energy dissipation per cycle $ \Delta E_{\mathrm{eddy}} $ can be calculated by averaging this energy loss within the cycle. 
By comparing this to the kinetic energy $ E_t = m A_z^2 \omega_z^2 / 2 $ where $m = \rho_{\mathrm{y}} \, 4/3 \pi a^3$ is the mass of the YIG sphere and $A_z$ is the amplitude of oscillation taken from the position of the YIG, the $Q$-factor $ Q_{\mathrm{eddy}} = 2 \pi E_t / (\Delta E_{\mathrm{eddy}}) $ is estimated. 


\begin{figure}[!b]
\begin{center}
\includegraphics[width= 6cm, clip]{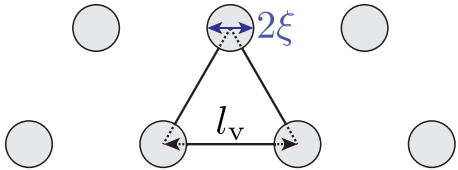}
\caption{Triangular vortex lattice of YBCO in mixed state, with lattice constant $ l_{\mathrm{v}} $, coherence length $\xi$. } 
\label{fig:fluxpin_YBCO}
\end{center}
\end{figure}

\section{Appendix I: Damping from YBCO in mixed state}

The YBCO in the vortex solid state is a mixed state where normal and superconducting regions coexist. 
If the external magnetic field is above $ \mathbf{B} \geq H_{\mathrm{c1}}$, vortices penetrate into the superconductor forming a triangular lattice configuration of lattice constant $ l_{\mathrm{v}} = 1.075 \sqrt{ \Phi / B_z } $, where $ \Phi = 2.068... \times 10^{-15} $ Wb is the magnetic flux quantum (Fig.~\ref{fig:fluxpin_YBCO})~\cite{Tinkham}.  
The vortex core is a normal region the size of the coherence length $\xi$, and a magnetic field penetrated region the size of $\lambda_{\mathrm{L}}$. 
Thus the volumetric ratio of normal region within a unit cell of the triangular lattice is 
\begin{equation}
\rho_{\mathrm{n}} 
= \frac{ \displaystyle\frac{\pi \xi^2}{2}}{ \displaystyle\frac{\sqrt{3}}{4} l_{\mathrm{v}}^2} 
= \frac{2 \pi \, \xi^2}{\sqrt{3} \, l_{\mathrm{v}}^2} 
= \frac{2 \pi}{1.156 \sqrt{3} \, \Phi} \, \xi^2 B_z .  
\tag{I1}
\end{equation}
Assuming eddy current damping occurs only in the normal regions and not in the superconducting regions, the eddy current damping of the mixed state can be estimated from the eddy current damping when the YBCO in a normal state $Q_{\mathrm{eddy, \, YBCO, \, n}}$ as 
\begin{equation}
Q_{\mathrm{eddy, \, YBCO}} 
= \frac{2 \pi E_t}{\rho_n \Delta E_{\mathrm{eddy}}} 
= \frac{ Q_{\mathrm{eddy, \, YBCO, \, n}}}{\rho_n}. 
\tag{I2}
\end{equation} 
Note that the vortex core size is much smaller than the skin depth of the ceramic YBCO in a normal state, and the actual $Q$-factor may be higher than estimated.

\section{Appendix J: Squeezed film damping estimation}

Isothermal squeezed film damping (SQFD) in atmospheric pressure is governed by both viscous and inertial effects of the air, which can be simulated by the nonlinear Reynolds equation~\cite{Bao-book}.
However, for high vacuum systems where the mean free path of gas molecules becomes much larger than the gap distance, the viscous flow model is no more valid, and the free molecular model has to be considered. 
Here we use the model proposed by Bao~\cite{Bao-RFmodel} who calculated the mechanical $Q$-factor of an oscillating plate with a neighboring surface using the energy transfer model to be
\begin{gather}
Q_{\mathrm{Sqfl}} 
= \frac{8 \sqrt{\pi} \rho_{\mathrm{y}} \omega_y }{\sqrt{2}} \frac{V}{\overline{l}^2} \frac{d_0}{L} \sqrt{\frac{RT}{M_m}} \frac{1}{P}
\tag{J1}
\end{gather}
with density $ \rho_{\mathrm{y}} = 5172 \, \mathrm{kg/m^3} $, $ V = 4/3 \pi a^3 $ the volume, $d_0$ the gap distance between the oscillating plate and neighboring surface, $\overline{l}$ the average travelling distance of a gas molecule within this gap, $L$ the peripheral length, gas constant $R = 8.31 \, \mathrm{m^2 \, kg \, s^{-2} \, K^{-1} \, mol^{-1}} $, temperature $T = 4$ K, pressure $ P = 10^{-5} $ Pa, and molar weight of air $ M_{\mathrm{m}} = 28.966 $ g/mol. 
Since SQFD is strongest against horizontal displacements, we assume the angular frequency is $ \omega_y = 2 \pi f_y $, and treat the superconductor as a cylinder, ignoring the slit for simplicity. 
In this case, $ d_0 = r - a $, $ L = 2 \pi a $, and 
\begin{gather}
\overline{l} 
= \frac{1}{2a} \int_{-a}^a \left( r - \sqrt{a^2 - z^2} \right)^2 \, dz 
= r^2 - \frac{2}{3} a^2 - \frac{\pi}{2} r a. 
\tag{J2}
\end{gather}
Thus the SQFD of a YIG sphere inside a cylinder is
\begin{gather}
Q_{\mathrm{sq}} = \frac{16 \rho_{\mathrm{y}}}{3} \sqrt{ \frac{R T}{M_{\mathrm{m}}} } \frac{a^2 (r-a)}{r^2 + 2/3 a^2 - \pi/2 a r} \frac{\omega}{P}.
\tag{J3}
\end{gather} 
Although the SQFD limited $Q$-factor is roughly an order lower than the vacuum limited $Q$-factor, $ Q_{\mathrm{sq}} > 10^{12} $ and is negligible at high enough vacuum $ P = 10^{-5}$ Pa compared to eddy current losses.

\section{Appendix K: Force noise from external magnetic field noise}

The magnetic force on the YIG sphere is given by the surface integration of the magnetic field energy 
\begin{gather}
\mathbf{F} 
= \int \int \frac{\mathbf{B}_n \cdot \mathbf{B}_n }{2 \, \mu_0} \, d \mathrm{S}. 
\tag{\ref{eq:horizontal_force}}
\end{gather}
where $\mathbf{B}_n$ is the normal component of the field to the surface. 
Here we consider the force fluctuations caused by the fluctuations in the current source $ I_{\mathrm{coil}} $ used to generate the external magnetic field $\mathbf{B}$. 
Since 
\begin{align*}
\frac{ \partial \mathbf{F}}{\partial I_{\mathrm{coil}} } 
&= \int \int \frac{\partial}{\partial I_{\mathrm{coil}} } \, \frac{ \mathbf{B}_n^2 }{2 \mu_0} \, d \mathrm{S} \\ 
&= \int \int \frac{1}{2 \mu_0} \, 2 \mathbf{B}_n \, \frac{\partial \mathbf{B}_n}{\partial I_{\mathrm{coil}} } \, d \mathrm{S}, 
\end{align*} 
the force fluctuation is 
\begin{equation*}
\delta F 
= \frac{ \partial \mathbf{F}}{\partial I_{\mathrm{coil}} } \, \delta I_{\mathrm{coil}} 
= \int \int \frac{1}{2 \mu_0} \, 2 \mathbf{B}_n \, \delta \mathbf{B}_n \, d \mathrm{S}. 
\end{equation*}
Since the magnetic field generated by a solenoid coil is proportional to the current applied $ \mathbf{B} \propto I_{\mathrm{coil}} $, 
$ \delta I_{\mathrm{coil}} / I_{\mathrm{coil}} = \delta \mathbf{B} / B $. 
In this case the force fluctuation is 
\begin{align*}
\delta F 
&= \int \int \frac{1}{2 \mu_0} \, 2 \mathbf{B}_n \, \left( \frac{\delta I_{\mathrm{coil}}}{I_{\mathrm{coil}}} \, \mathbf{B}_n \right) d \mathrm{S} \\
&= 2 \frac{\delta I_{\mathrm{coil}} }{I_{\mathrm{coil}}} \int \int \frac{ \mathbf{B}_n \cdot \mathbf{B}_n }{2 \mu_0} \, d \mathrm{S} \\
&= 2 \, \frac{\delta I_{\mathrm{coil}} }{I_{\mathrm{coil}}} \, F
\end{align*}
Since any external magnetic field fluctuation $\delta B$ can be modeled as an equivalent current fluctuation $\delta I_{\mathrm{coil}}$ of a solenoid coil, 
\begin{equation}
\frac{\delta F}{F} 
= 2 \, \frac{\delta I_{\mathrm{coil}}}{I_{\mathrm{coil}}}
= 2 \, \frac{\delta B}{B}. 
\tag{K1}
\end{equation}
This result matches the current fluctuation induced force fluctuation calculated using the harmonic perturbation study in COMSOL multiphysics.


\begin{thebibliography}{0}%
\makeatletter
\providecommand \@ifxundefined [1]{%
 \@ifx{#1\undefined}
}%
\providecommand \@ifnum [1]{%
 \ifnum #1\expandafter \@firstoftwo
 \else \expandafter \@secondoftwo
 \fi
}%
\providecommand \@ifx [1]{%
 \ifx #1\expandafter \@firstoftwo
 \else \expandafter \@secondoftwo
 \fi
}%
\providecommand \natexlab [1]{#1}%
\providecommand \enquote  [1]{``#1''}%
\providecommand \bibnamefont  [1]{#1}%
\providecommand \bibfnamefont [1]{#1}%
\providecommand \citenamefont [1]{#1}%
\providecommand \href@noop [0]{\@secondoftwo}%
\providecommand \href [0]{\begingroup \@sanitize@url \@href}%
\providecommand \@href[1]{\@@startlink{#1}\@@href}%
\providecommand \@@href[1]{\endgroup#1\@@endlink}%
\providecommand \@sanitize@url [0]{\catcode `\\12\catcode `\$12\catcode
  `\&12\catcode `\#12\catcode `\^12\catcode `\_12\catcode `\%12\relax}%
\providecommand \@@startlink[1]{}%
\providecommand \@@endlink[0]{}%
\providecommand \url  [0]{\begingroup\@sanitize@url \@url }%
\providecommand \@url [1]{\endgroup\@href {#1}{\urlprefix }}%
\providecommand \urlprefix  [0]{URL }%
\providecommand \Eprint [0]{\href }%
\providecommand \doibase [0]{http://dx.doi.org/}%
\providecommand \selectlanguage [0]{\@gobble}%
\providecommand \bibinfo  [0]{\@secondoftwo}%
\providecommand \bibfield  [0]{\@secondoftwo}%
\providecommand \translation [1]{[#1]}%
\providecommand \BibitemOpen [0]{}%
\providecommand \bibitemStop [0]{}%
\providecommand \bibitemNoStop [0]{.\EOS\space}%
\providecommand \EOS [0]{\spacefactor3000\relax}%
\providecommand \BibitemShut  [1]{\csname bibitem#1\endcsname}%
\let\auto@bib@innerbib\@empty
\end{thebibliography}%


\begin{thebibliography}{99}


\bibitem{Gonzalez-BallesteroRev}
C.\ Gonzalez-Ballestero, M.\ Aspelmeyer, L.\ Novotny, R.\ Quidant, and O.\ Romero-Isart, 
Science \textbf{374}, 168 (2021).

\bibitem{MaComPhys}
J.\ Ma, J.\ Qin, G.\ T.\ Campbell, G.\ Guccione, R.\ Lecamwasam, B.\ C.\ Buchler, and P.\ K.\ Lam, 
Communications Physics \textbf{3}, 197 (2020). 

\bibitem{QinOptica}
J.\ Qin, G.\ Guccione, J.\ Ma, C.\ Gu, R.\ Lecamwasam, B.\ C.\ Buchler, and P.\ K.\ Lam, 
Optica \textbf{9}, 924 (2022). 

\bibitem{RomeroPRL12}
O.\ Romero-Isart, L.\ Clemente, C.\ Navau, A.\ Sanchez, and J.\ I.\ Cirac,
\PRL{109}{147205} (2012).

\bibitem{CirioPRL}
M.\ Cirio, G.\ K.\ Brennen, and J.\ Twamley, 
\PRL{109}{147206} (2012). 

\bibitem{DigiacomoNanoMat}
L.\ Digiacomo, E.\ Quagliarini, B.\ Marmiroli, B.\ Sartori, G.\ Perini, M.\ Papi, A.\ L.\ Capriotti, C.\ M.\ Montone, A.\ Cerrato, G.\ Caracciolo and D.\ Pozzi, 
Nanomaterials \textbf{12}, 2376 (2022). 

\bibitem{GieselerPRL}
J.\ Gieseler, A.\ Kabcenell, E.\ Rosenfeld, J.\ D.\ Schaefer, A.\ Safira, M.\ J.\ A.\ Schuetz, C.\ Gonzalez-Ballestero, C.\ C.\ Rusconi, O.\ Romero-Isart, and M.\ D.\ Lukin, 
\PRL{124}{163604} (2020). 

\bibitem{WangPRAp}
T.\ Wang, S.\ Lourette, S.\ R.\ O’Kelley, M.\ Kayci, Y.\ B.\ Band, Derek F.\ Jackson Kimball, A.\ O.\ Sushkov, and D.\ Budker, 
Phys. Rev. Applied \textbf{11}, 044041 (2019).

\bibitem{HoferArxiv}
J.\ Hofer, G.\ Higgins, H.\ Huebl, O.\ F.\ Kieler, R.\ Kleiner, D.\ Koelle, P.\ Schmidt, J.\ Slater, M.\ Trupke, K.\ Uhl, T.\ Weimann, W.\ Wieczorek, F.\ Wulschner, and M.\ Aspelmeyer,
arXiv:2211.06289 [quant-ph]. 

\bibitem{LewandowskiPRAp}
C.\ W.\ Lewandowski, T.\ D.\ Knowles, Z.\ B.\ Etienne, and B.\ D’Urso, 
Phys. Rev. Applied \textbf{15}, 014050 (2021).

\bibitem{LengPRAp}
Y.\ Leng, R.\ Li, X.\ Kong, H.\ Xie, D.\ Zheng, P.\ Yin, F.\ Xiong, T.\ Wu, C.-K.\ Duan, Y.\ Du, Z.\ Q.\ Yin, P.\ Huang, and J.\ Du,
Physical Review Applied \textbf{15} 024061 (2021)

\bibitem{VinantePRap}
A.\ Vinante, P.\ Falferi, G.\ Gasbarri, A.\ Setter, C.\ Timberlake, and H.\ Ulbricht,
Physical Review Applied \textbf{13} 064027 (2020).

\bibitem{LatorreIEEE}
M.\ Gutierrez Latorre, A.\ Paradkar, D.\ Hambraeus, G.\ Higgins, and W.\ Wieczorek,
IEEE Transactions on Applied Superconductivity \textbf{32}, 1800305 (2022). 

\bibitem{LatorrePRAp}
M.\ Gutierrez Latorre, G.\ Higgins, A.\ Paradkar, T.\ Bauch, and W.\ Wieczorek,
Physical Review Applied \textbf{19} 054047 (2023).

\bibitem{SchuckSciA}
M.\ Schuck, D.\ Steinert, T.\ Nussbaumer, and J.\ W.\ Kolar, 
Science Advances \textbf{4}, e1701519 (2018). 

\bibitem{XiongPRAp}
F.\ Xiong, P.\ Yin, T.\ Wu, H.\ Xie, R.\ Li, Y.\ Leng, Y.\ Li, C.\ Duan, X.\ Kong, P.\ Huang, and J.\ Du, 
Phys. Rev. Applied \textbf{16}, L011003 (2021).

\bibitem{RautIEEE}
N.\ K.\ Raut, J.\ Miller, J.\ Pate, R.\ Chiao, and J.\ E.\ Sharping, 
IEEE Transactions on Applied Superconductivity \textbf{31}, 1500204 (2021). 

\bibitem{JiangAPL}
X.\ Jiang, J.\ Rudge, and M.\ Hosseini, 
 Appl. Phys. Lett. \textbf{116}, 244103 (2020). 

\bibitem{NakajimaPRA}
R.\ Nakashima, 
Physics Letters A \textbf{384} 126592 (2020). 

\bibitem{ChenAS}
X.\ Chen, S.\ K.\ Ammu, K.\ Masania, P.\ G.\ Steeneken, and F.\ Alijani, 
Advanced Science \textbf{9}, 2203619 (2022). 

\bibitem{RomagnoliArXiv} 
P.\ Romagnoli, R.\ Lecamwasam, S.\ Tian, J.\ E.\ Downes, and J. Twamley, 
arXiv:2211.08764v1 [physics.app-ph]. 

\bibitem{Prat-CampsPRL}
J.\ Prat-Camps, C. Teo, C. C. Rusconi, W. Wieczorek, and O. Romero-Isart, 
Phys. Rev. Applied \textbf{8}, 034002 (2017).

\bibitem{Goodkind}
J.\ M.\ Goodkind, 
Review of Scientific Instruments \textbf{70}, 4131 (1999). 

\bibitem{KimballPRL}
D.\ F.\ Jackson Kimball, A.\ O. Sushkov, and D.\ Budker,
\PRL{116}{190801} (2016).

\bibitem{SebersonJOSAB}
T.\ Seberson, P.\ Ju, J.\ Ahn, J.\ Bang, T.\ Li, and F.\ Robicheaux, 
Journal of the Optical Society of America B \textbf{37} 3714 (2020). 

\bibitem{TimberlakeAPL}
C.\ Timberlake, G.\ Gasbarri, A.\ Vinante, A.\ Setter, and H.\ Ulbricht, 
Appl. Phys. Lett. \textbf{115}, 224101 (2019). 

\bibitem{BallesteroPRL}
C.\ Gonzalez-Ballestero, J.\ Gieseler, and O.\ Romero-Isart,
\PRL{124}{093602} (2020).

\bibitem{KaniPRLcool}
A.\ Kani, B.\ Sarma, and J.\ Twamley, 
\PRL{128}{013602} (2022).


\bibitem{FadeevQST}
P.\ Fadeev, C.\ Timberlake, T.\ Wang, A.\ Vinante, Y.\ B Band, D.\ Budker, A.\ O Sushkov, H.\ Ulbricht, and Derek F.\ Jackson Kimball, 
Quantum Sci. Technol. \textbf{6} 024006 (2021). 

\bibitem{FadeevPRD}
P.\ Fadeev, T.\ Wang, Y.\ B.\ Band, D.\ Budker, P.\ W.\ Graham, A.\ O.\ Sushkov, and D.\ F.\ J.\ Kimball, 
Phys.\ Rev.\ D \textbf{103} 044056, (2021). 

\bibitem{GrimsmoPRX}
A.\ L.\ Grimsmo, J.\ Combes, and B.\ Q.\ Baragiola, 
Phys.\ Rev.\ X \textbf{10} 011058, (2020). 


\bibitem{Earnshaw}
S.\ Earnshaw,
Transactions of the Cambridge Philosophical Society \textbf{7}, 97 (1842). 

\bibitem{MoonWiley}
F. C. Moon,  
Wiley (1994).

\bibitem{HardingNASA}
J.T. Harding, 
JPL Technical Report \textbf{30}, 806 (1965).


\bibitem{SpencerPRL}
E.\ G.\ Spencer, R.\ C.\ LeCraw, and A.\ M.\ Clogston,
\PRL{3}{32} (1959).

\bibitem{ZhangPRL}
X.\ Zhang, C.\-L.\ Zou, L.\ Jiang, and H.\ X.\ Tang, 
\PRL{113}{156401} (2014). 

\bibitem{TabuchiPRL}
Y.\ Tabuchi, S.\ Ishino, T.\ Ishikawa, R.\ Yamazaki, K.\ Usami, and Y.\ Nakamura, 
\articletitle{Hybridizing Ferromagnetic Magnons and Microwave Photons in the Quantum Limit}
Phys. Rev. Lett. \textbf{113}, 083603 (2014).

\bibitem{BallesteroPRB}
C.\ Gonzalez-Ballestero, D.\ H\"ummer, J.\ Gieseler, and O.\ Romero-Isart,
Phys.\ Rev.\ B \textbf{101} 125404, (2020). 



\bibitem{YIGmu-init}
A.\ Siblini, I.\ Khalil, J.\ P.\ Chatelon, J.\ J.\ Rousseau, 
Advanced Materials Research \textbf{324}, 290 (2011). 

\bibitem{YIGdielectric}
Deltronic Crystal Industries Product Description

\bibitem{KomissarenkoNanoMat}
F.\ Komissarenko, G.\ Zograf, S.\ Makarov, M.\ Petrov, and I.\ Mukhin, 
Nanomaterials \textbf{10}, 1306 (2020). 

\bibitem{BaeRevIn}
S.\ Bae, Y.\ Tan, A.\ P.\ Zhuravel, L.\ Zhang, S.\ Zeng, Y.\ Liu, T.\ A.\ Lograsso, A.\ T.\ Venkatesan, and S.\ M.\ Anlage, 
Rev. Sci. Instrum. \textbf{90}, 043901 (2019). 

\bibitem{HakkiIEEE}
B.\ W.\ Hakki, and P.\ D.\ Coleman, 
IEEE Trans. Microwave Theory Tech. \textbf{8}, 402 (1960).

\bibitem{HashimotoSci}
K.\ Hashimoto, K.\ Cho, T.\ Shibauchi, S.\ Kasahara, Y.\ Mizukami, R.\ Katsumata, Y.\ Tsuruhara, T.\ Terashima, H.\ Ikeda, M.\ A.\ Tanatar, H.\ Kitano, N.\ Salovich, R.\ W.\ Giannetta, P.\ Walmsley, A.\ Carrington, R.\ Prozorov, and Y.\ Matsuda, 
Science \textbf{336}, 6088 (2012). 


\bibitem{NbHc1}
R.\ A.\ French, 
Cryogenics \textbf{8}, 301 (1968). 

\bibitem{NbHc2}
S.\ J.\ Williamson, 
Phys. Rev. B \textbf{2}, 3545 (1970). 

\bibitem{YBCOHc1}
R.\ Liang, P.\ Dosanjh, D.\ A.\ Bonn, W.\ N.\ Hardy, and A.\ J.\ Berlinsky,
Phys.\ Rev.\ B \textbf{50}, 4212 (1994).

\bibitem{YBCOHvs}
G.\ Grissonnanche, O.\ Cyr-Choini\'ere, F.\ Lalibert\'e, \textit{et al.}, 
Nat.\ Commun.\ \textbf{5}, 3280 (2014). 

\bibitem{NbLondon}
B.\ W.\ Maxfield and W.\ L.\ McLean, 
Phys. Rev. \textbf{139}, A1515 (1965). 

\bibitem{YBCOLondon}
M.\ E.\ McHenry and R.\ A.\ Sutton,
Progress in Material Science \textbf{38}, 159 (1994). 



\bibitem{Tinkham}
M.\ Tinkham,  
``Introduction to superconductivity", 
Dover, 2004. 

\bibitem{NamburiSup}
D.\ K.\ Namburi, Y.\ Shi, and D.\ A.\ Cardwell, 
Supercond. Sci. Technol. \textbf{34}, 053002 (2021). 



\bibitem{Bao-RFmodel}
M.\ Bao, H.\ Yang, H.\ Yin and Y.\ Sun, 
J. Micromech. Microeng. \textbf{12}, 341 (2002). 

\bibitem{SchmoranzerCryo}
D.\ Schmoranzer, A.\ Luck, E.\ Collin, and A.\ Fefferman, 
Cryogenics \textbf{98}, 102 (2019). 

\bibitem{Fleischer-RevIn}
S.\ M.\ Fleischer, M.\ P.\ Ross, K.\ Venkateswara, C.\ A.\ Hagedorn, E.\ A.\ Shaw, E.\ Swanson, B.\ R.\ Heckel, and J.\ H.\ Gundlach, 
Rev. Sci. Instrum. \textbf{93}, 064505 (2022).


\bibitem{LIGO-budget}
A.\ Buikema, \textit{et. al.}, 
Phys.\ Rev.\ D \textbf{102}, 062003 (2020). 

\bibitem{SMC11}
For example, SMC11 Puy Mary, from Sisyph.  

\bibitem{ILL2019} 
D.\ Wurm, D.\ H.\ Beck, T.\ Chupp, S.\ Degenkolb, K.\ Fierlinger, P.\ Fierlinger, H.\ Filter, S.\ Ivanov, C.\ Klau, M.\ Kreuz, E.\ Leli\'evre-Berna, T.\ Lins, J.\ Meichelböck, T.\ Neulinger, R.\ Paddock, F.\ R\"ohrer, M.\ Rosner, A.\ P.\ Serebrov, J.\ T.\ Singh, R.\ Stoepler, S.\ Stuiber, M.\ Sturm, B.\ Taubenheim, X.\ Tonon, M.\ Tucker, M.\ van der Grinten and O.\ Zimmer, 
\articletitle{The PanEDM neutron electric dipole moment experiment at the ILL}
EPJ Web Conf. \textbf{219}, 02006 (2019). 

\bibitem{RusconiPRA}
C.\ C.\ Rusconi, M.\ J.\ A.\ Schuetz, J.\ Gieseler, M.\ D.\ Lukin, and O.\ Romero-Isart, 
Phys.\ Rev.\ A \textbf{100} 022343, (2019). 

\bibitem{WernsdorferSci}
W.\ Wernsdorfer, and R.\ Sessoli, 
Science \textbf{284}, 133 (1999). 

\bibitem{KeshtgarPRB}
H.\ Keshtgar, S.\ Streib, A.\ Kamra, Y.\ M.\ Blanter, and G.\ E.\ W.\ Bauer, 
Phys.\ Rev.\ B \textbf{95} 134447, (2017). 

\bibitem{RusconiPRL17}
C.\ C.\ Rusconi, V.\ P\"ochhacker, K.\ Kustura, J.\ I.\ Cirac, and O.\ Romero-Isart,
\PRL{119}{167202} (2017).

\bibitem{RusconiPRB16}
C.\ C.\ Rusconi and O.\ Romero-Isart, 
Phys.\ Rev.\ B \textbf{93} 054427, (2016). 

\bibitem{KaniPRL22}
A.\ Kani, F.\ Quijandr\'ia, and J.\ Twamley, 
\PRL{129}{257201} (2022).

\bibitem{MaPRL}
Y.\ Ma, K.\ E.\ Khosla, B.\ A.\ Stickler, and M.\ S.\ Kim,
\PRL{125}{053604} (2020).

\bibitem{DelordNature}
T.\ Delord, P.\ Huillery, L.\ Nicolas and G.\ H\'etet, 
Nature \textbf{580}, 56 (2020).

\bibitem{HuilleryPRB}
P.\ Huillery, T.\ Delord, L.\ Nicolas, M.\ Van Den Bossche, M.\ Perdriat, and G.\ H\'etet, 
Phys.\ Rev.\ B \textbf{101} 134415 , (2020). 

\bibitem{PerdriatRev}
M.\ Perdriat, C.\ Pellet-Mary, P.\ Huillery, L.\ Rondin, and G.\ H\'etet, 
Micromachines \textbf{12}, 65 (2021). 

\bibitem{YIGlevexp}
M.\ Fuwa, R.\ Sakagami, and T.\ Tamegai, 
arXiv:2306.13917 [quant-ph] (2023). 



\bibitem{Earnshaw}
S.\ Earnshaw,
Transactions of the Cambridge Philosophical Society \textbf{7}, 97 (1842). 

\bibitem{Bao-book}
M.\ Bao, 
Elsevier, Amsterdam, 2005 (Chapter 3).



\end{thebibliography}
\end{document}